# Magnetic Lifting of BIC Symmetry Protection Enables Nonreciprocal Beam Steering and Angular Information Encoding

Chengrui Zhang[1], Zijing Xianyu[1], Chufan Zhou[2]*, Qi Jiang[1], Shengxuan Si[1], Wanli Xue[1], Ji Wu[1], Haoming Zhong[1], Jiaqi Zhang[1], Hanqi He[1], Edoardo Charbon[2]*, Zhifeng Zhang[3,4*], Zhiyu Wang[1]*

[1]School of Mechanical Science and Engineering, Huazhong University of Science and Technology, Wuhan, China.

E-mail: wangzhiyu023@hust.edu.cn

[2]Advanced Quantum Architecture Laboratory (AQUA), Ecole Polytechnique Federale de Lausanne (EPFL), 2002 Neuchatel, Switzerland.

E-mail: chufan.zhou@epfl.ch; edoardo.charbon@epfl.ch

[3]National Laboratory of Solid State Microstructures, Key Laboratory of Intelligent Optical Sensing and Manipulation, Nanjing University, 210093, Nanjing, China.

[4]College of Engineering and Applied Sciences, and Collaborative Innovation Center of Advanced Microstructures, Nanjing University, 210093, Nanjing, China.

E-mail: zfzhang@nju.edu.cn

## Abstract

Symmetry-protected bound states in the continuum (SP-BICs) offer exceptional optical confinement but remain nonradiative. Here, we demonstrate that magnetically lifting their protecting symmetry enables nonreciprocal beam steering with intrinsic angular information encoding. In a magneto-optical photonic crystal slab, an in-plane magnetic field converts an SP-BIC into a radiative quasi-BIC and, together with vertical dielectric asymmetry, continuously displaces the associated band extremum away from Γ. The steered beam exhibits two independent locking relations: its wavelength is locked to the elevation angle, while its polarization orientation is locked to the azimuthal angle. These relations embed complementary spectral and polarization information into the scanning beam, providing additional sensing channels for multidimensional LiDAR. Moreover, the nonreciprocal dispersion, $\omega(+k) \neq \omega(-k)$, supports resonant emission at $+k$, while back-reflected light at the same frequency is off resonance at $-k$, opening a route to intrinsic self-isolation and reduced optical feedback. These findings establish a unified framework connecting magnetic beam steering, angular information encoding and nonreciprocal feedback suppression, with prospects for compact reconfigurable light sources and integrated optical engines.

Bound states in the continuum (BICs), originally conceived in quantum mechanics and now ubiquitous across photonic architectures (*1-3*), have become central to modern optics owing to their theoretically infinite quality factors (*4*) and underlying topological nature (*5,6*). Symmetry-protected BICs (SP-BICs), whose topological protection arises from in-plane rotational symmetry, are robust against symmetry-preserving geometric perturbations (*7-10*). Their polarization singularity guarantees a perfectly nonradiative state at the Γ point, i.e., along the surface normal (*11-15*). For practical light sources such as lasers, radiation can only be extracted from states near Γ; however, this inevitably degrades beam collimation and leaves the far-field polarization ill-defined.

To enhance beam collimation and engineer the far-field polarization, breaking in-plane symmetry to convert an SP-BIC into a quasi-BIC is routinely employed (*16-22*). As illustrated in **Fig. 1A**, deforming a rotationally symmetric unit-cell void pattern yields a quasi-BIC with designed polarization at normal emission (*23-26*). However, in reciprocal systems, the band dispersion is necessarily an even function of the wave vector, constrained by time-reversal symmetry and lattice periodicity (*25-27*). The band extremum therefore remains pinned at the Γ point, where the vanishing group velocity provides high-quality optical feedback that favors lasing strictly along the surface normal. Alternatively, as shown in **Fig. 1B**, Brillouin-zone folding using supercell configurations (*6,9,28-30*) creates multiple folded band extrema (marked as $\Gamma_0$, $\Gamma_{+1}$, $\Gamma_{-1}$) that enable oblique emission with designed polarization. Yet, all these geometry-based strategies only yield fixed, discrete emission angles, leaving continuous and dynamic beam steering around the Γ point an unresolved challenge.

Here, we break this constraint by moving to a nonreciprocal platform. As shown in **Fig. 1C**, using a magneto-optical photonic crystal (MO-PhC) slab under an in-plane magnetic field ($H_x$), we lift the protective symmetry and convert an SP-BIC into a quasi-BIC. In the presence of cladding–substrate refractive-index asymmetry, the same field shifts the zero-group-velocity band extremum away from Γ, enabling continuous, nonmechanical tuning of the emission elevation angle ($\varphi_y$). The resulting nonreciprocal dispersion, ($\omega(+k_y) \neq \omega(-k_y)$), supports resonant emission at $+k_y$, while back-reflected light at the same frequency is off resonance at $-k_y$, opening a route to intrinsic self-isolation and reduced optical feedback. Beyond steering the beam, magnetic control establishes two intrinsic locking relations: the linear polarization orientation is locked to the azimuthal angle, while the emission frequency is locked to the elevation angle. These relations encode complementary spectral and polarization information into the steered beam, providing additional sensing channels for multidimensional LiDAR.

The MO-PhC slab features a periodicity of $a$ = 850 nm, a thickness of $t$ = 200 nm, and a square-hole side length of $L$ = 460 nm, with cladding and substrate indices $n_1$ and $n_2$, as shown in **Fig. 2A**. An in-plane magnetic field $H_x$ induces a magnetization $M_x$ in the MO material, which gives rise to off-diagonal permittivity components $\delta_x$ described by the tensor

$$\varepsilon = \begin{pmatrix} \varepsilon_d & 0 & 0 \\ 0 & \varepsilon_d & i\delta_x \\ 0 & -i\delta_x & \varepsilon_d \end{pmatrix} \quad (1)$$

where $\varepsilon_d$ is the isotropic dielectric permittivity, taken as $\varepsilon_d = 6.25$, a value representative of lossless magneto-optical thin films such as bismuth iron garnet (BIG), Cobalt Ferrite (CFO), and cerium-doped YIG (Ce:YIG). We first examine the nonmagnetic SP-BIC by setting $\delta_x = 0$ and $n_1 = n_2$. The photonic band structure in momentum space (left panel of **Fig. 2B**) is symmetric about Γ, with the band extremum located exactly at Γ and an infinite $Q$-factor. This mode is an SP-BIC protected by in-plane $C_4$ rotational symmetry and remains robust against symmetry-preserving geometric perturbations (**Fig. S1**). The electric field inside the slab exhibits a tightly confined dipolar profile with no outgoing radiation (right panel of **Fig. 2B**), confirming its nonradiative nature. In the far-field, this manifests as a characteristic annular far-field radiation-intensity pattern centered at Γ with a central void (inset of **Fig. 2B**), consistent with previous reports (*4,31*).

To explore the effect of magnetization, we set $\delta_x = 0.5$, a value that corresponds to magnetic fields of several hundred mT in magneto-optical materials (*33-35*), while keeping $n_1 = n_2$. As shown in the left panel of **Fig. 2C**, the band structure remains symmetric about Γ, but the $Q$-factor drops to a finite value on the order of $10^5$. Since $\delta_x$ is introduced without material absorption, this $Q$ reduction arises entirely from the radiative channel opened due to in-plane rotational symmetry broken by $M_x$. As shown in the right panel of **Fig. 2C**, the eigenmode profile at Γ now displays outgoing radiative waves above and below the slab, with the original dipolar profile slightly twisted. Correspondingly, the far-field radiation-intensity pattern turns into a bright spot centered at Γ (inset of **Fig. 2C**), indicating that the magnetically induced quasi-BIC fills the central void and angular divergence of the original SP-BIC. This mechanism thus provides a means to tune SP-BICs into quasi-BICs. Notably, an out-of-plane magnetization $M_z$ alone cannot lift the protective symmetry and therefore leaves the BIC intact, in agreement with earlier work (*3*).

We further introduce a refractive-index asymmetry between the cladding and the substrate ($n_1 \neq n_2$), which breaks the vertical symmetry. Under $M_x$, this shifts the band extremum away from Γ toward $-k_y$ (the left panel of **Fig. 2D**). The far-field radiation-intensity pattern (inset of **Fig. 2D**) reveals a bright spot displaced from the center, corresponding to a beam deflection toward $-y$. This oblique radiation is indicated by the tilted arrow in the right panel of **Fig. 2D**. This demonstrates that the in-plane magnetization, together with the vertical index asymmetry, can convert the SP-BIC into a quasi-BIC while steering the emission. Meanwhile, this nonreciprocal band shift ($\omega(k_y) \neq \omega(-k_y)$) suppresses the coupling of back-reflected light to the emitting mode, providing an intrinsic optical-isolation capability. The corresponding far-field polarization textures for the three configurations in **Figs. 2B–D** are presented in **Fig. S2**, illustrating how the polarization singularity evolves during the SP-BIC–to–quasi-BIC transition. Remarkably, the deflection occurs exclusively along $k_y$, with no shift along $k_x$. This behavior raises two fundamental questions: (i) why does the vertical refractive-index asymmetry unpin the band extremum from Γ, and (ii) why does an in-plane magnetization along $x$ produce a deflection only along $y$? The answers lie in the interplay between magnetically broken symmetries and residual spatial symmetries of the structure.

The first question is resolved by examining the $C_{2x}$ rotational symmetry of the MO-PhC slab about the $x$-axis. In the index-symmetric case of **Fig. 2C**, the structure retains $C_{2x}$ symmetry even under $M_x$. As sketched in **Fig. 2E**, if the band extremum were shifted to $k_y$ (corresponding to an elevation angle $\varphi_y$), $C_{2x}$ symmetry would require a counterpart at $-k_y$ ($-\varphi_y$). These two symmetry-required extrema can merge into a single extremum only if they coincide at $k_y=0$, thereby pinning the extremum to Γ. When the cladding and substrate become dissimilar (i.e., the case of **Fig. 2D**), this $C_{2x}$ constraint is lifted, allowing the magneto-optically induced effective gauge field to shift the extremum away from Γ.

The second question is resolved by the $\sigma_x$ mirror symmetry of the MO-PhC slab. As illustrated in **Fig. 2F**, $M_x$ is an axial vector oriented along $x$ and is therefore invariant under $\sigma_x$; hence the entire system remains symmetric under this mirror operation. Since $\sigma_x$ maps a deflection angle $\varphi_x$ to $-\varphi_x$, a mirror-invariant extremum can exist only when $\varphi_x \equiv 0$. Consequently, the band extremum is strictly confined to move along $k_y$. This symmetry-enforced constraint provides a simple and robust rule for magneto-optical beam steering and can be extended to other $C_6$-symmetric MO-PhC structures that retain the corresponding mirror symmetry (see **Fig. S3**).

Geometry imposes that the band extremum shifts orthogonally to the applied magnetic field. Interestingly, however, for a fixed magnetic field along $x$-axis, the shift along $k_y$ can have opposite signs (i.e., toward $+k_y$ or $-k_y$), depending strongly on the modal structure (i.e., the electric-field distribution in the MO-PhC slab). As shown in **Fig. 3A**, to illustrate how band-structure and modal properties govern these steering characteristics, we examine two modes, Mode B and Mode C, both of which are SP-BICs at the Γ point with infinite $Q$-factor in the absence of a magnetic field. When $\delta_x$ is set to 0.5, their band extrema shift in opposite $k_y$ directions (**Figs. 3B** and **3C**). To understand how the magnetic field controls the band-extremum shift, we analyze the interaction between the modal electric field and the magnetically induced off-diagonal permittivity terms,

$$\Delta\boldsymbol{\varepsilon}_{\boldsymbol{MO}} = \delta_x \begin{pmatrix} 0 & 0 & 0 \\ 0 & 0 & i \\ 0 & -i & 0 \end{pmatrix}. \tag{2}$$

The frequency correction of the photonic band can be expressed as

$$\frac{\Delta\omega^{MO}}{\omega} = -\frac{1}{2}\frac{\int \boldsymbol{E}^* \cdot \Delta\varepsilon_{MO} \cdot \boldsymbol{E} dV}{\int \boldsymbol{E}^* \cdot \varepsilon_0 \cdot \boldsymbol{E} dV} = \frac{\delta_x \int \mathrm{Im}\left(E_y^* E_z\right) dV}{\int \boldsymbol{E}^* \boldsymbol{\varepsilon}_0 \boldsymbol{E} dV} = \delta_x S_x^{MO}. \tag{3}$$

Here, $S_x^{MO}$ denotes the ratio of the two volume integrals and characterizes the magneto-optical influence on the band. $\Delta\omega^{MO}$, $\omega$ and $S_x^{MO}$ are functions of $k_y$. Since $\omega$ is much larger than the magneto-optical correction $\Delta\omega^{MO}$ , we may replace ω by $\omega_0$ when considering $\Delta\omega^{MO}$. Near $k_y = 0$, $S_x^{MO}$ is linear in $k_y$, so the correction takes the form

$$\Delta\omega^{MO}\left(k_y\right) = \delta_x \omega\left(k_y\right) S_x^{MO}\left(k_y\right) \approx \delta_x \omega_0 \frac{\partial S_x^{MO}}{\partial k_y} k_y. \tag{4}$$

In the absence of a magnetic field, reciprocity constrains the band dispersion to be an even function of $k_y$:

$$\omega\left(k_y\right) \approx \omega_0 + \alpha k_y^2 + O\left(k_y^4\right). \tag{5}$$

With magnetization, the total dispersion becomes

$$\omega\left(k_y\right) = \omega^0\left(k_y\right) + \Delta\omega^{MO}\left(k_y\right) \approx \omega_0 + \alpha k_y^2 + \delta_x \omega_0 \frac{\partial S_x^{MO}}{\partial k_y} k_y. \tag{6}$$

Thus, the magnetic field effectively adds a linear term to the quadratic band, shifting the zero-group-velocity band extremum to $k_y^* = \left(-\delta_x \omega_0 / 2\alpha\right)\left(\partial S_x^{MO} / \partial k_y\right)$ . This expression provides a direct design rule for magnetic beam steering: the desired steering direction ($\pm k_y$) can be determined by engineering the relative signs of the band curvature $\alpha$ and the magneto-optical overlap gradient $\partial S_x^{MO} / \partial k_y$. At a fixed magnetization, the steering magnitude scales inversely with $|\alpha|$ and linearly with $|\partial S_x^{MO} / \partial k_y|$. Accordingly, flattening the photonic band (reducing |$\alpha$|) or enhancing the near-field overlap gradient ($|\partial S_x^{MO} / \partial k_y|$) provides a direct route to amplify the deflection angle for a given mode.

In the two cases considered here, both bands open downward ($\alpha$<0), so the sign of $\alpha$ is identical. The spatial profiles of $\mathrm{Im}\left(E_y^* E_z\right)$ in **Figs. 3F** and **3G** yield opposite signs of $\partial S_x^{MO} / \partial k_y$ at Γ for Modes B and C. Accordingly, the band extrema shift in opposite directions, precisely as observed. The same rule is confirmed for all modes in the relevant spectral range (**Table S1** and **Fig. S4**), establishing a direct mathematical mapping from the near-field distribution to the direction and magnitude of the band-extremum shift.

As shown in **Figs. 3D and 3E**, the *Q*-factor maximum nearly co-migrates with the band extremum under magnetic perturbation. This near-coincidence ensures that the zero-group-velocity point benefits from the high *Q*-factor, which is favorable for developing beam-steering functionality in threshold light sources (such as lasers and polariton condensates). However, comparing **Figs. 3D and 3E** with **Figs. 3B and 3C** reveals that the two extrema shift by different amounts in momentum space. Although the *Q*-factor is related to the imaginary part of the eigenfrequency (radiative loss), it does not simply follow its real counterpart, i.e., the photonic band, $\omega\left(k_y\right)$. As detailed in **Note S3**, besides the band shift, the actual local radiative loss also influences the evolution of the *Q*-factor maximum in a complicated manner. Consequently, the *Q*-factor maximum does not always follow the band extremum (examples where they shift in opposite directions are shown in **Fig. S4**).

As the band extremum shifts in momentum space, its frequency changes and the originally singular polarization becomes well-defined. **Fig. 4A** presents a comprehensive schematic of the steered emission properties, mapped onto a reference hemisphere centered at the MO-PhC slab. The emission direction is parameterized by the elevation angle $\varphi$ and the azimuthal angle $\theta$, which are controlled by the strength and the orientation of the applied in-plane magnetic field, respectively. Beyond these geometric correspondences, the emission frequency and polarization can also be mapped onto the hemisphere: each latitudinal ring ($\varphi$) corresponds to a unique emission frequency (iso-frequency contours labeled $\omega_0$, $\omega_1$, $\omega_2$, …), and each meridional line ($\theta$) is rigidly tied to a specific linear polarization direction $\theta_p$.

To decode the relation between $\varphi$ and $\omega$, we first plot in **Fig. 4B** the momentum-space photonic bands for increasing magnetization strength (parametrized by $\delta_x$ in Eq. 1). For the mode identical to Mode A, as $\delta_x$ increases, the band extremum shifts along $-k_y$ while its frequency also changes simultaneously. Associating the in-plane wave-vector magnitude $k_{||}$ at the band extremum with the elevation angle $\varphi$, we obtain the $\varphi$–$\omega$ relation shown in **Fig. 4C**, where the data points correspond to successive increments of $\delta_x$ by 0.5. The emission frequency increases monotonically with $\varphi$ across the simulated angular range, following an approximately quadratic trajectory. This locking between frequency $\omega$ and elevation angle $\varphi$ provides an extra dimension of information in beam steering: different elevation channels carry distinct spectral signatures, allowing the receiver to identify the emission angle through frequency-resolved detection and to reconstruct the corresponding angular information in momentum space. Although the intrinsic elevation-angle tuning demonstrated here spans 0-4.5°, the system-level angular range could be extended to tens of degrees using suitably designed external angular-magnification or metasurface-assisted optics (*36-38*).

In addition, the in-plane magnetic field simultaneously controls the far-field polarization associated with the displaced band extremum. Using the full gyrotropic permittivity tensor

$$\varepsilon = \begin{pmatrix} \varepsilon_d & i\delta_z & -i\delta_y \\ -i\delta_z & \varepsilon_d & i\delta_x \\ i\delta_y & -i\delta_x & \varepsilon_d \end{pmatrix}, \qquad (7)$$

we map the band frequency overlaid by the far-field polarization in **Fig. 4D** for an in-plane magnetic field with $H_x = H_y$ and zero $H_z$ ($\delta_x = \delta_y = 0.5$ and $\delta_z = 0$). The band extremum appears at $(k_x, k_y) \approx (0.006\pi/a, -0.006\pi/a)$, which in real space corresponds to an emission direction perpendicular to the applied in-plane magnetic field. Remarkably, the polarization remains perfectly aligned with the direction of the band-extremum shift, i.e., along $k_x = -k_y$ (indicated by different shades of red). This "polarization-invariant zone" is always pinned to that shift direction regardless of the field strength, so that a well-defined linear polarization is always imparted to the band extremum (see **Fig. S5** in the Supplementary Material). The first row of **Fig. 4F** illustrates this one-to-one correspondence by extracting the linear polarization state at various $\theta$. As $\theta$ is swept by rotating the in-plane magnetic field direction, the polarization orientation angle $\theta_p$ follows synchronously. This establishes a robust locking between $\theta_p$ and $\theta$ that is independent of the magnetic-field-strength-governed $\varphi$–$\omega$ locking. Operationally, during spatial beam scanning, the structure intrinsically maintains a predictable polarization state without any extra control.

We further find that this $\theta_p$-$\theta$ locking can be lifted or erased, by applying an additional perpendicular magnetic field $H_z$, which converts the linear polarization into circular polarization. As shown in the successive rows of **Fig. 4F**, increasing $\delta_z$ causes all linear polarization states to become gradually elliptical and eventually purely circular at a specific field strength, corresponding to a C point. **Fig. 4E** shows that, starting from the in-plane magnetized case, adding a given $\delta_z$ endows the band extremum with pure circular polarization. This polarization transition originates from the magnetically induced coupling between two orthogonal polarization channels. The out-of-plane magnetization introduces off-diagonal components $\pm i\delta_z$ in the permittivity tensor (i.e., $\varepsilon_{xy}$ and $\varepsilon_{yx}$), which modify the relative phase between the two radiated polarization components. As a result, the initially locked linear polarization gradually acquires a phase difference approaching $\pm\pi/2$, giving rise to circularly polarized emission. The detailed mechanism by which $\delta_z$ converts linear polarization into circular polarization has been discussed in previous studies (*34,39*). This process mainly reshapes the polarization state rather than the mode confinement, and the $Q$-factor remains nearly unchanged under this weak perturbation. This not only lifts the initial linear-polarization locking but, more importantly, enables the deflected beam to carry pure circular polarization, which can be exploited to probe planar chiral surface properties, thereby providing complementary information beyond the spatial coordinates obtained from beam steering. More generally, this work provides a universal picture of the evolution of a lossless MO-PhC slab under an applied magnetic field. The sensitivity of both beam steering and polarization conversion to the off-diagonal components of the permittivity tensor can vary strongly with the modal field distribution. Moreover, alternative geometries, such as MO-cylindrical pillar arrays, achieve circular polarization with a much smaller $\delta_z$, as shown in **Fig. S6**, while following the same underlying magneto-optical interaction mechanism.

In summary, we have demonstrated that an in-plane magnetic field applied to a magneto-optical photonic structure converts SP-BICs into quasi-BICs. This conversion is accompanied by a shift of the band extremum orthogonal to the applied field, thereby enabling beam steering with angle-locked emission frequency and polarization state. Different directions carry distinct spectral and polarization signatures, allowing the receiver to identify or reconstruct the corresponding information in momentum space. This capability could enable faster, more accurate LiDAR-based 3D imaging and chiral surface discrimination. Beyond LiDAR, magnetic beam steering enables fine control of the emission angle and, through the nonreciprocal band shift ($\omega(k_y) \neq \omega(-k_y)$), inherently suppresses the coupling of back-reflected light to the emitting mode without requiring a separate optical isolator. Such multifunctionality, supported by our analytical design rule for tailoring the steering direction and magnitude through modal engineering, may facilitate highly integrated miniature on-chip optical engines.

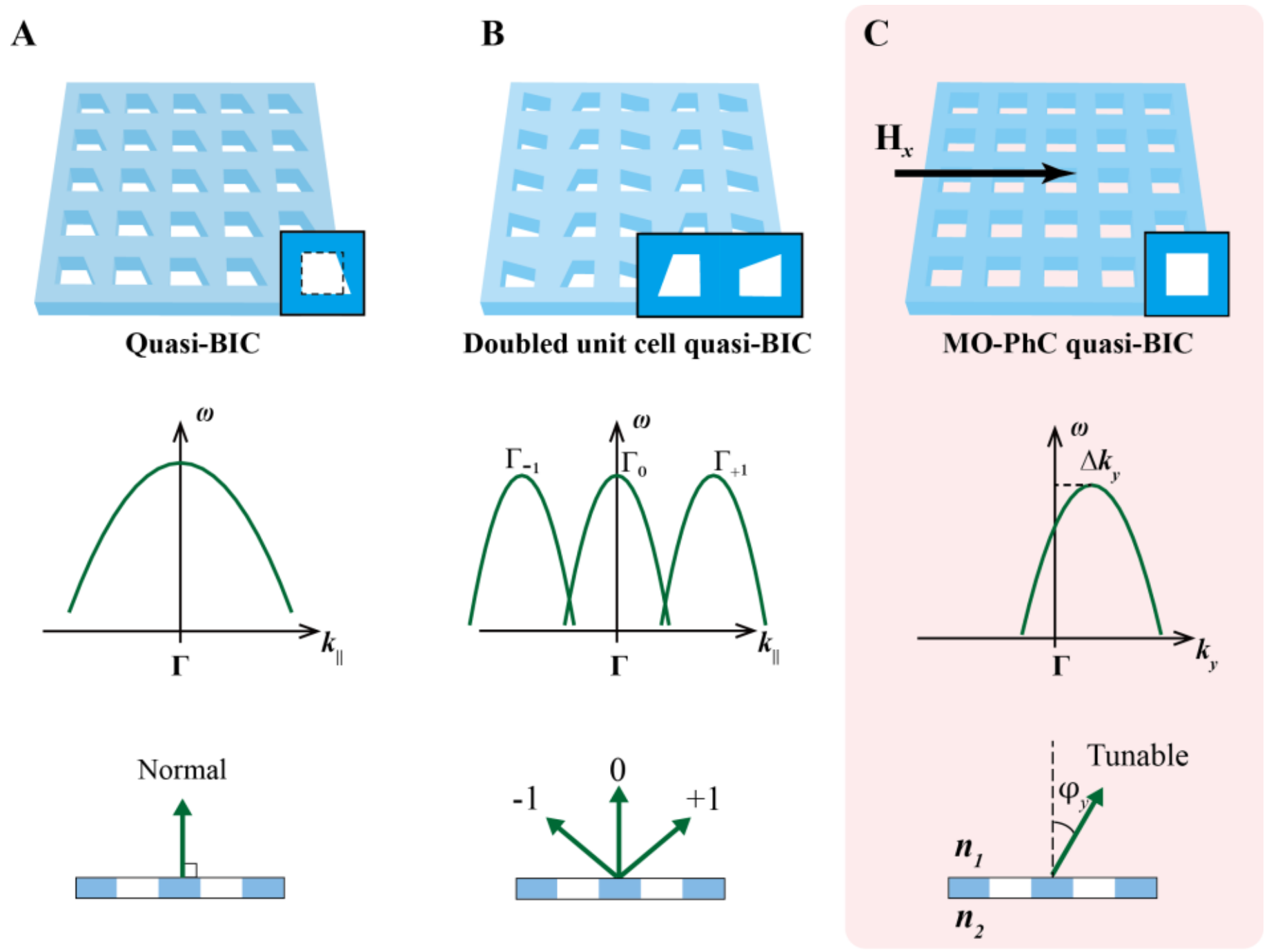

**Fig. 1. Optical signatures of quasi-BICs obtained via distinct SP-BIC-lifting strategies.**

(A) A quasi-BIC induced by geometric symmetry reduction: the band extremum remains pinned at Γ by reciprocity, restricting emission to the surface normal. (B) A doubled unit cell quasi-BIC: Brillouin-zone folding introduces discrete valleys, allowing oblique emission only at fixed orders. (C) The MO-PhC quasi-BIC concept: in-plane magnetization (induced by $H_x$) breaks the protective rotational symmetry, and the nonreciprocal system allows the band extremum to deviate from Γ, enabling continuously tunable oblique emission. Notably, a magnetic field applied along the in-plane $x$ direction deflects the band exclusively along $k_y$, where $y$ denotes the orthogonal in-plane axis.

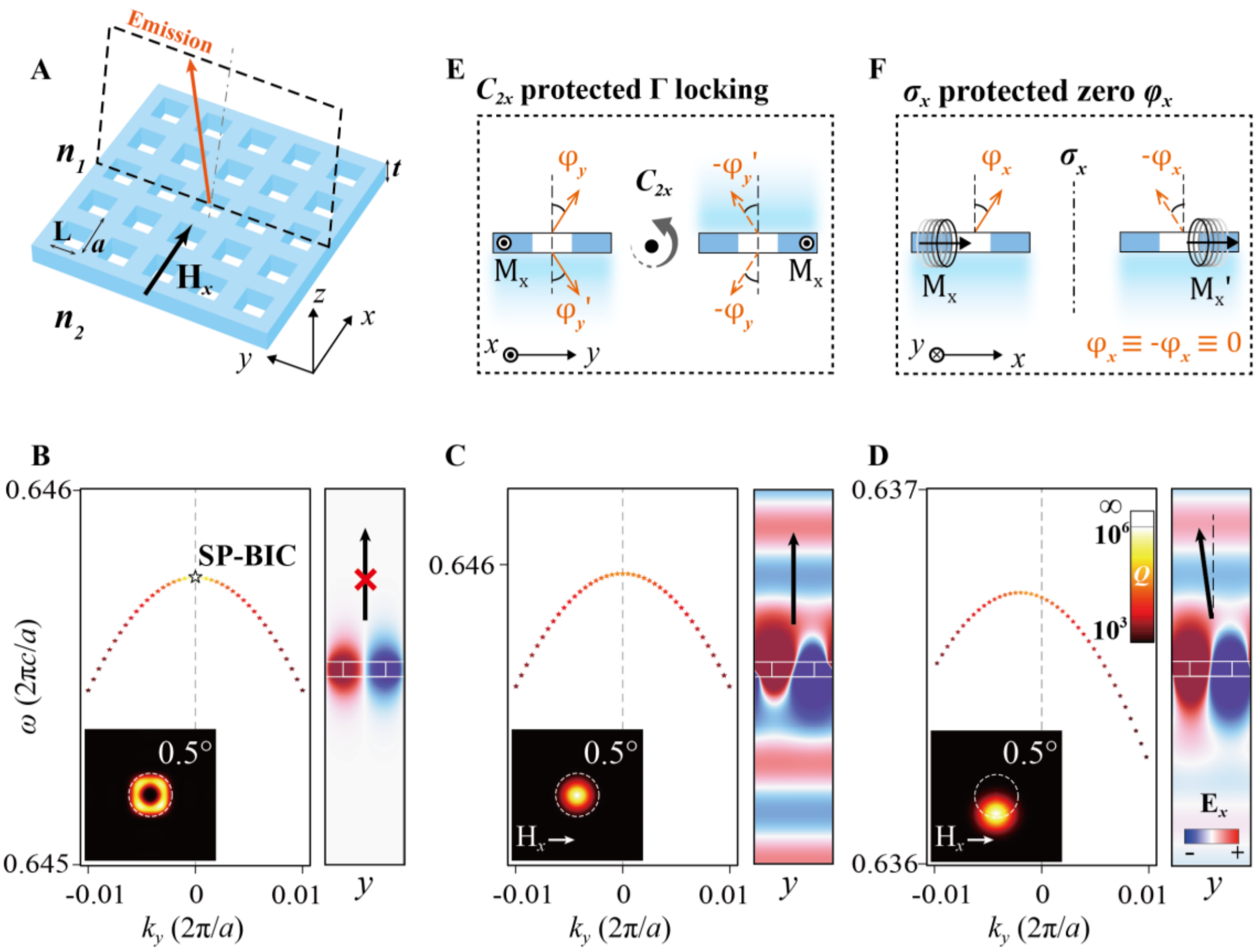

**Fig. 2. Magnetically induced SP-BIC lifting and beam steering.**

(A) Schematic of the MO-PhC slab (period $a$ = 850 nm, thickness $t$ = 200 nm, square hole side $L$ = 460 nm) with cladding and substrate indices $n_1$, $n_2$, and in-plane magnetic field $H_x$. The orange arrow illustrates the magnetically steered oblique emission, while the dashed plane indicates the y−z steering plane perpendicular to $H_x$. (B) The simulated results for the MO-PhC slab with $n_1 = n_2 = 1$ under zero magnetic field. Photonic bands (left) color coded by $Q$-factor, showing an infinite $Q$ at Γ (the SP-BIC). Inset: far-field radiation-intensity pattern with a central void; the white dashed circle indicates a wave vector range corresponding to a 0.5° emission angle in air (same scale in C and D). The side-view $E_x$ profile presents a left–right-symmetry dipole pattern, and no outgoing radiation, confirming its BIC character. (C) Applying $H_x$ ($n_1 = n_2 = 1$) lifts the SP-BIC into a quasi-BIC with finite $Q$ at Γ; the far-field radiation intensity fills the central void, and the mode profile loses left–right-symmetry, showing outgoing radiative waves. (D) Adding index asymmetry ($n_1 = 1$, $n_2 = 1.2$) under the same $H_x$ shifts the band extremum and far-field radiation-intensity maximum along $-k_y$ (orthogonal to $H_x$). The $E_x$ field is sampled at the photonic band extremum, corresponding to oblique emission (the arrow is a guide). The $Q$ and $E_x$ color bars in (D) apply to (B)–(D). (E) The Γ-locking in (C) is attributed to the $C_{2x}$ rotational symmetry, which pins the band extremum at Γ in the index-symmetric environment, as illustrated in (E). The absence of any $x$-deflection in (D) (shift only along $y$) follows from the $\sigma_x$ mirror symmetry: $\sigma_x$ enforces $\varphi_x \equiv -\varphi_x \equiv 0$, resulting in a deflection purely along $k_y$, as sketched in (F).

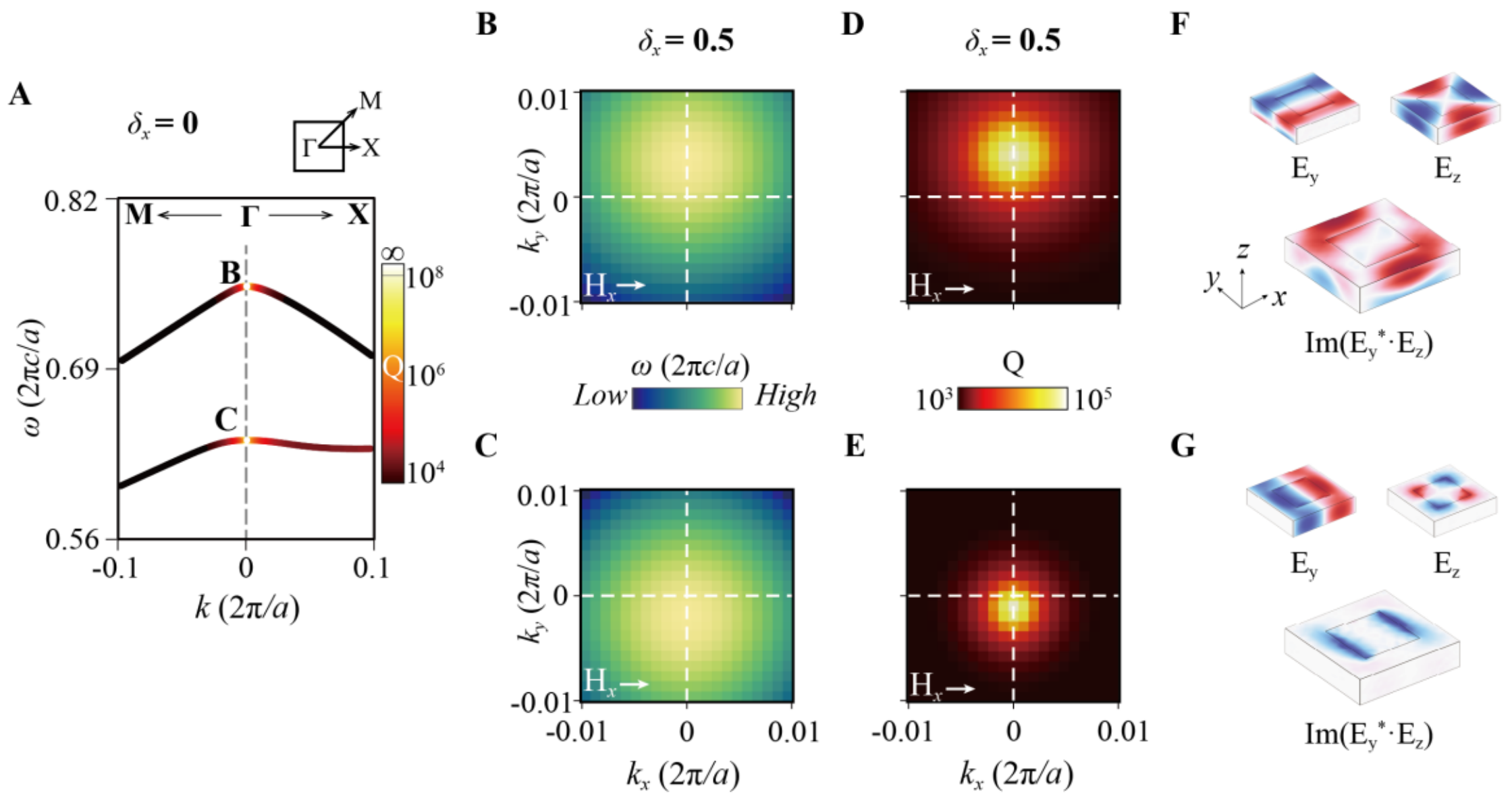

**Fig. 3. Mode-dependent magneto-optical response governed by the virtual field.**

(A) Photonic band dispersions of two selected parent SP-BIC modes in the vertically symmetric and unmagnetized MO-PhC slab ($\delta_x$ = 0). Both modes exhibit band extrema at the Γ point. Mode C corresponds to the SP-BIC mode used in Figs. 2 and 4. (B and C) Momentum-space frequency distributions after introducing the cladding–substrate index asymmetry ($n_1$ = 1, $n_2$ = 1.2) and an in-plane magnetization $M_x$ ($\delta_x$ = 0.5). The band extrema of the two modes shift along opposite $k_y$ directions. (D and E) Corresponding momentum-space $Q$-factor distributions under the same magnetic perturbation. The $Q$-factor maxima approximately follow the displaced band extrema. (F and G) Modal origins of the opposite magneto-optical responses. The electric-field components $E_y$ and $E_z$ together with the local magneto-optical overlap term $\mathrm{Im}(E_y^* E_z)$, are shown for Modes B and C. The normalized integration of this overlap term over the magneto-optical region gives opposite signs of $\partial S_x^{MO} / \partial k_y$ at Γ, leading to opposite photonic band shift in momentum-space.

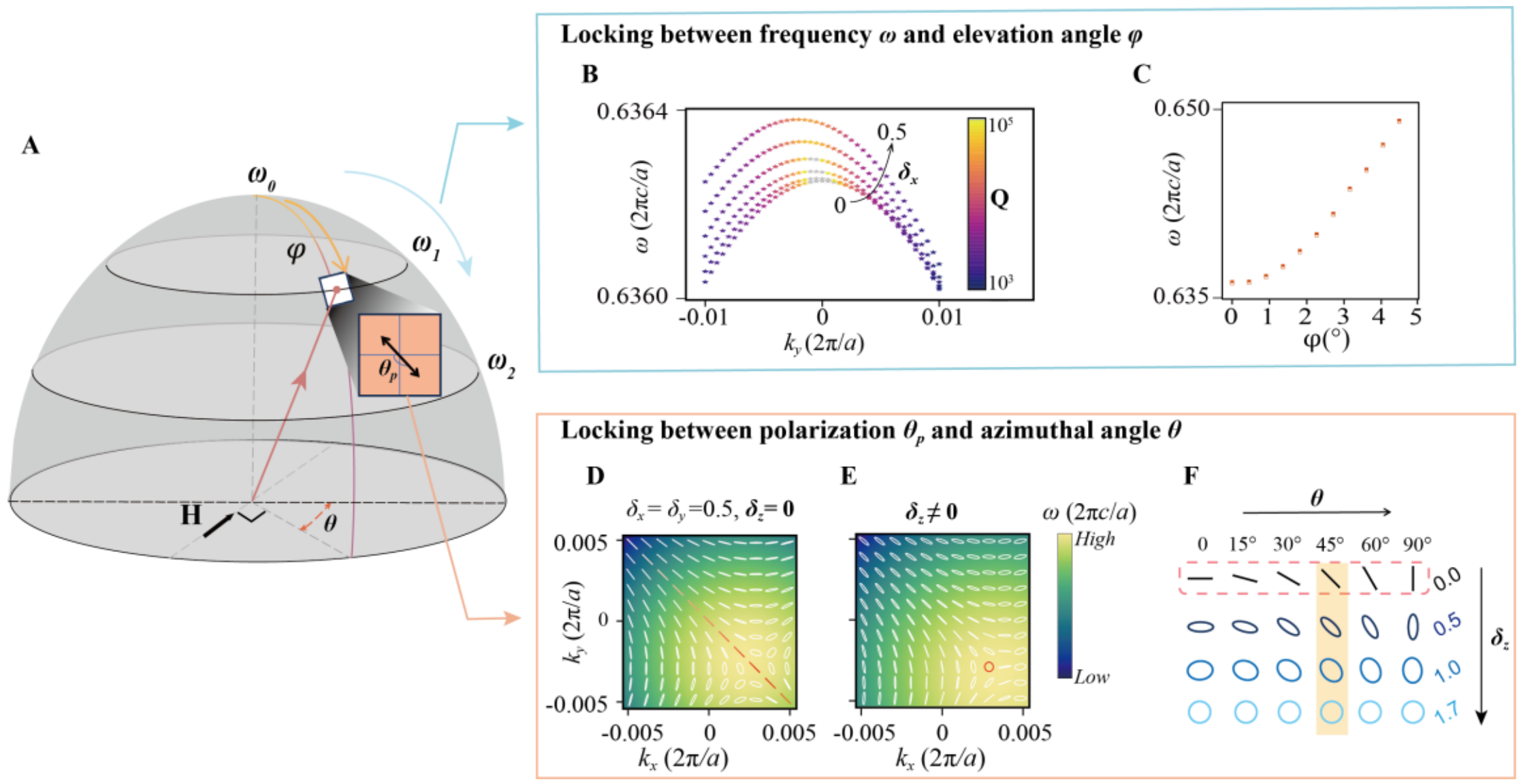

**Fig. 4. Magnetically Steered Emission with Angle-Locked Frequency and Polarization.**

(A) Comprehensive schematic of the steered emission properties depicted on an emission reference hemisphere centered at the MO-PhC slab. Longitude and latitude correspond to the azimuthal angle $\theta$ and elevation angle $\varphi$, respectively; the emission frequencies are labeled $\omega_0$, $\omega_1$, $\omega_2$…, and the linear-polarization orientation angle is defined as $\theta_p$. (B) Momentum-space photonic bands for increasing $\delta_x$ (only $H_x$ is applied, $\delta_y = \delta_z = 0$), color-coded by the $Q$-factor. (C) Band-extremum frequency as a function of $\varphi$, by associating the in-plane wave-vector magnitude $k_{||}$ at the band extremum with the elevation angle $\varphi$. A one-to-one locking between $\omega$ and $\varphi$ is evident. (D) Momentum-space band frequency and overlaid polarization states for an in-plane magnetic field with $\delta_x = \delta_y = 0.5$, $\delta_z = 0$. Along the direction of the band-extremum shift ($k_x = -k_y$), the polarization remains linear (indicated in red). (E) Same as (D) but for a three-dimensional magnetic field with $\delta_x = \delta_y = 0.5$, $\delta_z = 1.7$. A purely circular polarization state (marked in red) appears at the band extremum. (F) The array of polarization states at the band extremum (corresponding to the far-field emission) is shown for varying $\theta$ (horizontal) and increasing $\delta_z$ (vertical). In the first row (with in-plane magnetic field only), each $\theta$ is locked to a distinct linear polarization orientation angle $\theta_p$, demonstrating the $\theta_p$-$\theta$ locking. As $\delta_z$ increases, all linear states become gradually elliptical and eventually purely circular at $\delta_z = 1.7$. $H_z$ can lift the linear polarization locking and control the degree of circular polarization.

**Acknowledgments**
We thank Professor Takuo Tanemura of The University of Tokyo for insightful discussions and valuable contributions to the development of this manuscript. **Funding:** supported by the Overseas Outstanding Youth Foundation of the National Natural Science Foundation of China (0214100245) **Author contributions:** C.Z. and Z.W. conceived the study and analyzed the data. C.Z. performed the numerical simulations and wrote the original draft. Z.X. performed the theoretical validation. Z.W., Z.Z and E.C. reviewed and edited the manuscript. All authors discussed the results and contributed to the final manuscript. **Competing interests:** The authors declare that they have no competing interests. **Data and materials availability:** All data needed to evaluate the conclusions in the paper are present in the paper and/or the Supplementary Materials.

**Supplementary Information**

## Note S1. Numerical methods

All numerical simulations were performed using COMSOL Multiphysics. A single unit cell of the magneto-optical photonic crystal slab was modeled with Bloch periodic boundary conditions along the x and y directions and perfectly matched layers along the z direction. The magneto-optical response was incorporated through the gyrotropic permittivity tensor defined in Eq. 1 of the main text. The photonic bands, eigenmode profiles, and Q-factor distributions were obtained using the eigenfrequency solver by sweeping the in-plane Bloch wave vector ($k_x$, $k_y$) around the Γ point. The Q factor was calculated from Q=Re($\tilde{\omega}$)/2|Im($\tilde{\omega}$)|, where Re($\tilde{\omega}$)and Im($\tilde{\omega}$)denote the real and imaginary parts of the eigenfrequency of the simulated eigenmode.

The far-field polarization textures were extracted from the corresponding single-unit-cell eigenmode fields. Only the far-field radiation-intensity patterns shown in the insets of Figs. 2B–D were calculated separately using a finite 15×15-unit-cell array. Material absorption was neglected throughout the simulations.

## Note S2. Topological Charge of the Parent SP-BIC and Evolution of the Far-Field Polarization Texture

The topological charge q describes the winding number of polarization axis of leaky modes around BICs, which is defined as

$$q = \frac{1}{2\pi}\oint_L d\,k_{||} \cdot \nabla_{k_{||}} \varphi(k_{||}). \qquad (S1)$$

Here, $L$ denotes a closed loop enclosing the BIC, and $\varphi(k_{||})$ represents the azimuthal angle of the major axis of the polarization state. For this mode, the topological charge is $q$ = -1.

## Note S3. Perturbative Analysis of the Frequency- and Q-Factor-Extremum Shifts

For magnetization along the (+x) direction, the following sign convention is adopted:

$$\varepsilon = \begin{pmatrix} \varepsilon_d & 0 & 0 \\ 0 & \varepsilon_d & i\delta_x \\ 0 & -i\delta_x & \varepsilon_d \end{pmatrix} = \varepsilon_0 + \Delta\varepsilon_{MO}. \tag{S2}$$

To quantify the cladding–substrate index asymmetry, we introduce the fixed structural parameter $\eta = (n_2 - n_1)/(n_2 + n_1)$. Unless otherwise specified, $\eta$ is kept fixed throughout the following perturbative analysis.

Under the weak magneto-optical coupling and nondispersive approximations, the first-order frequency correction can be written as

$$\frac{\Delta\omega^{MO}}{\omega} = -\frac{1}{2}\frac{\int_{MO} \boldsymbol{E}^* \cdot \Delta\varepsilon_{MO} \cdot \boldsymbol{E} dV}{\int_{cell} \boldsymbol{E}^* \cdot \varepsilon_0 \cdot \boldsymbol{E}\, dV}. \tag{S3}$$

For the above dielectric tensor,

$$\boldsymbol{E}^* \cdot \Delta\varepsilon_{MO} \cdot \boldsymbol{E} = -2\delta_x Im\left(E_y^* E_z\right). \tag{S4}$$

We therefore define the normalized MO-active transverse-spin overlap as

$$S_x^{MO} = \frac{2\int_{MO} Im(E_y^* E_z) dV}{\int_{cell} \boldsymbol{E}^* \cdot \varepsilon_0 \cdot \boldsymbol{E}\, dV}. \tag{S5}$$

Substituting Eqs. (S4) and (S5) into Eq. (S3) gives

$$\frac{\Delta\omega^{MO}}{\omega} = \frac{\delta_x}{2} S_x^{MO}. \tag{S6}$$

The integrand in Eq. (S5) is related to the x component of the electric-field spin density because

$$[Im(\boldsymbol{E}^* \times \boldsymbol{E})]_x = 2 Im\left(E_y^* E_z\right). \tag{S7}$$

Thus, $S_x^{MO}$ quantifies the part of the modal transverse optical spin that spatially overlaps with the MO material and couples to $M_x$.

**Shift of the band extremum**

The reciprocal background dispersion remains even in $k_y$ and can be expanded near the Γ point as

$$\omega^0(k_y) = \omega_0 + \alpha k_y^2 + O\left(k_y^4\right), \tag{S8}$$

where

$$\alpha = \frac{1}{2}\frac{\partial^2 \omega^0}{\partial k_y^2}\Big|_{k_y=0} \tag{S9}$$

is the local band-curvature coefficient. A local frequency maximum at Γ corresponds to $\alpha < 0$, whereas a local minimum corresponds to $\alpha > 0$.

Because $S_x^{MO}$ is odd in $k_y$ and vanishes in the vertically symmetric limit, its leading-order expansion can be written as

$$S_x^{MO}(k_y) = s\eta k_y + O(k_y^3), \quad (S10)$$

where $s\eta$ is an asymmetry-normalized, mode-dependent coefficient defined through

$$s\eta = \frac{\partial S_x^{MO}}{\partial k_y}|_{k_y=0}. \quad (S11)$$

The perturbed frequency is written as

$$\omega(k_y) = \omega^0(k_y) + \Delta\omega^{MO}(k_y). \quad (S12)$$

Substituting Eq (S10) into Eq. (S6) gives, to leading order,

$$\Delta\omega^{MO}(k_y) = \frac{\delta_x}{2}\omega^0(k_y)S_x^{MO}(k_y) \approx \frac{\omega_0\delta_x s\eta}{2}k_y. \quad (S13)$$

The magneto-optically perturbed local dispersion is therefore

$$\omega(k_y) = \omega_0 + \alpha k_y^2 + \frac{\omega_0\delta_x s\eta}{2}k_y. \quad (S14)$$

For consistency with the compact notation used in the main text, we define the effective magneto-optical coupling coefficient as

$$g = \frac{\omega_0 s\eta}{2}. \quad (S15)$$

So that Eq. (S14) becomes

$$\omega(k_y) = \omega_0 + \alpha k_y^2 + g\delta_x k_y. \quad (S16)$$

The band extremum is obtained from the zero-group-velocity condition

$$\frac{\partial\omega}{\partial k_y}|_{k_y=k_\omega^*} = 0. \quad (S17)$$

Using Eq. (S16), its position is approximately

$$k_\omega^* \approx -\frac{g\delta_x}{2\alpha}. \quad (S18)$$

Therefore, the direction criterion is

$$sgn(k_\omega^*) = -sgn\left(\frac{g\delta_x}{\alpha}\right) = -sgn\left(\frac{\delta_x s\eta}{\alpha}\right). \quad (S19)$$

Thus, the band-extremum displacement vanishes when either the magneto-optical coupling is absent or the vertical structure is index symmetric. Reversing either the magnetization or the cladding–substrate index contrast reverses the predicted displacement direction.

**Shift of the Q-factor maximum**

The analysis above describes the displacement of the frequency extremum, corresponding to the real part of the eigenfrequency. To examine the accompanying evolution of the Q factor, we write the complex eigenfrequency as

$$\tilde{\omega}(k_y) = \omega(k_y) - i\gamma(k_y), \qquad (S20).$$

$$Q^{-1}(k_y) = \frac{2\gamma(k_y)}{\omega(k_y)}. \qquad (S21)$$

Here, $\gamma$ denotes the decay rate. Because material absorption is neglected in our simulations， $\gamma$ is entirely determined by radiative loss.

The same broken symmetries also permit an odd-in-$k_y$ contribution to the radiative-loss dispersion. Near Γ, the latter can therefore be expanded as

$$\gamma(k_y) \approx \gamma_0 + \beta k_y^2 + b\eta\delta_x k_y, \qquad (S22)$$

where $\gamma_0$ is the decay rate at Γ, $\beta$ describes the local curvature of the radiative-loss distribution, and the coefficient $b$ characterizes the leading mixed response of the radiative decay rate to the Bloch wave vector and magneto-optical coupling,

$$b\eta = \frac{\partial^2\gamma}{\partial k_y \partial \delta_x}|_{k_y=\delta_x=0}. \qquad (S23)$$

Substituting Eqs. (S16) and (S22) into Eq. (S21), expanding the denominator about $\omega_0$ and retaining only the lowest nonvanishing terms up to second order in the joint ($k_y$, $\delta_x$) expansion—namely $k_y^2$ and $\delta_x k_y$—while neglecting higher-order contributions such as $k_y^4$, $\delta_x k_y^3$, and $\delta_x^2 k_y^2$, gives

$$Q^{-1}(k_y) \approx Q_0^{-1} + \alpha_Q k_y^2 + g_Q \delta_x k_y, \qquad (S24)$$

with

$$Q_0^{-1} = \frac{2\gamma_0}{\omega_0}, \quad (S25)$$

$$\alpha_Q = \frac{2(\omega_0\beta - \gamma_0\alpha)}{{\omega_0}^2}, \quad (S26)$$

And

$$g_Q = \frac{2(\omega_0 b\eta - \gamma_0 g)}{{\omega_0}^2}. \quad (S27)$$

Because both $b\eta$ and $g$ vanish when $\eta = 0$, the effective odd coefficient $g_Q$ also vanishes in the vertically symmetric limit.

Maximizing Q is equivalent to minimizing $Q^{-1}$, the Q-factor maximum is obtained from

$$\frac{\partial Q^{-1}}{\partial k_y}|_{k_y=k_Q^*} = 0. \quad (S28)$$

Its position is therefore approximately

$$k_Q^* = -\frac{g_Q\delta_x}{2\alpha_Q}, \quad (S29)$$

or, equivalently,

$$k_Q^* = -\frac{\delta_x(\omega_0 b\eta - \gamma_0 g)}{2(\omega_0\beta - \gamma_0\alpha)}. \quad (S30)$$

The corresponding direction criterion is

$$sgn(k_Q^*) = -sgn\left(\frac{g_Q\delta_x}{\alpha_Q}\right). \quad (S31)$$

Thus, the frequency extremum and Q-factor maximum approximately co-migrate when

$$\frac{g}{\alpha} \approx \frac{g_Q}{\alpha_Q}. \quad (S32)$$

This relation is not imposed by symmetry: $g, \alpha, g_Q$ and $\alpha_Q$ are all mode-dependent quantities.

It can be found that the Q-factor maximum does not necessarily coincide exactly with either the frequency extremum or the radiative-loss minimum. For Modes B and C shown in Fig. 3, the calculated radiative-loss minima remain close to the displaced frequency extrema, and the corresponding Q-factor maxima consequently migrate approximately together with them. This near-coincidence is consistent with an approximately common local momentum shift of the real and imaginary parts of the complex eigenfrequency. However, it remains a mode-dependent response rather than a rigid translation enforced by symmetry; when Eq. (S32) is not satisfied, the frequency and Q-factor extrema can separate or even shift in opposite directions.

Figs. S4D-F present the momentum-space Q-factor distributions and the migration of the corresponding Q extrema for the additional representative modes shown in Figs. S4A–C, whereas Figs. S4G-I display their frequency distributions and the associated migration of the band extrema. The arrows indicate the respective migration directions along $k_y$. For the modes shown in Figs. S4A and S4B, the Q extrema exhibit co-migration with the corresponding band extrema. In contrast, for the mode shown in Fig. S4C, the Q extremum and the band extremum shift in opposite directions, and no such co-migration occurs. These results further demonstrate that the migration of the Q extrema is mode dependent and does not necessarily follow that of the band extrema.

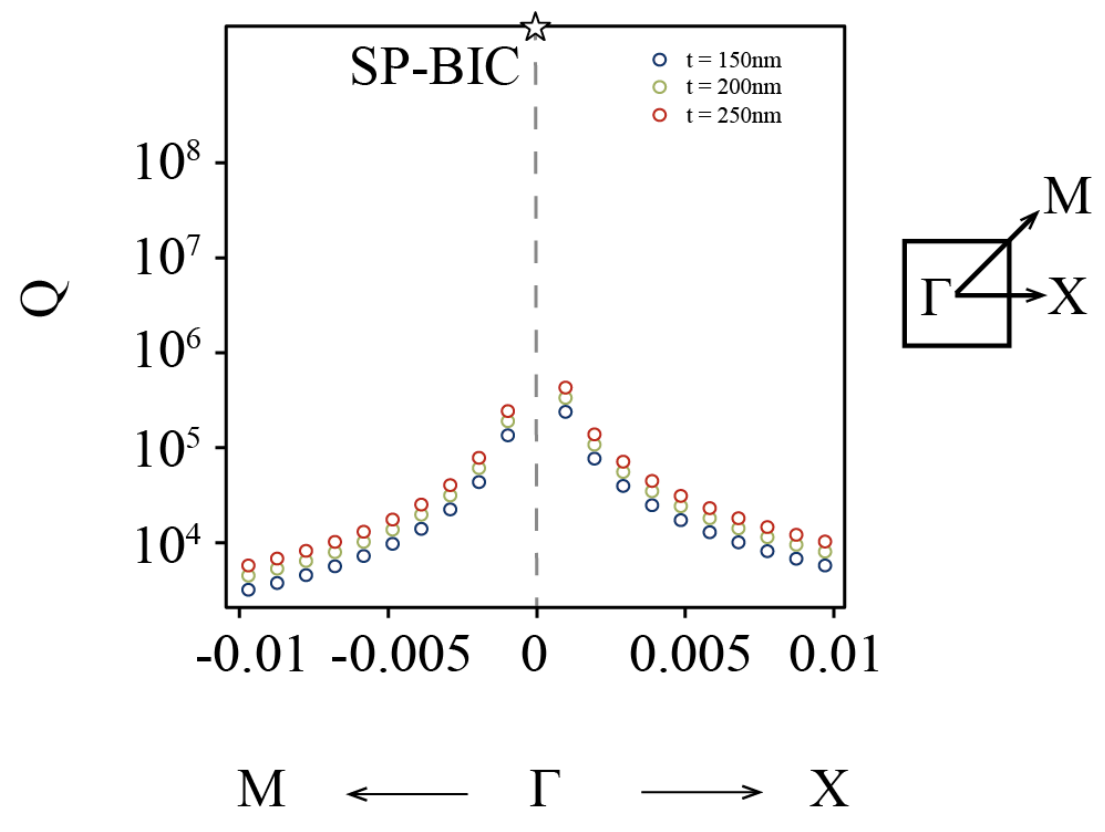

**Fig. S1 | Robustness of the SP-BIC against symmetry-preserving thickness variations.**

Q factors calculated along the M–Γ–X path for slab thicknesses of $t$ = 150, 200, and 250 nm. In all three cases, the Q factor exhibits a numerically divergent peak at Γ, demonstrating that the SP-BIC remains pinned at the Γ point when the in-plane $C_4$ rotational symmetry is preserved.

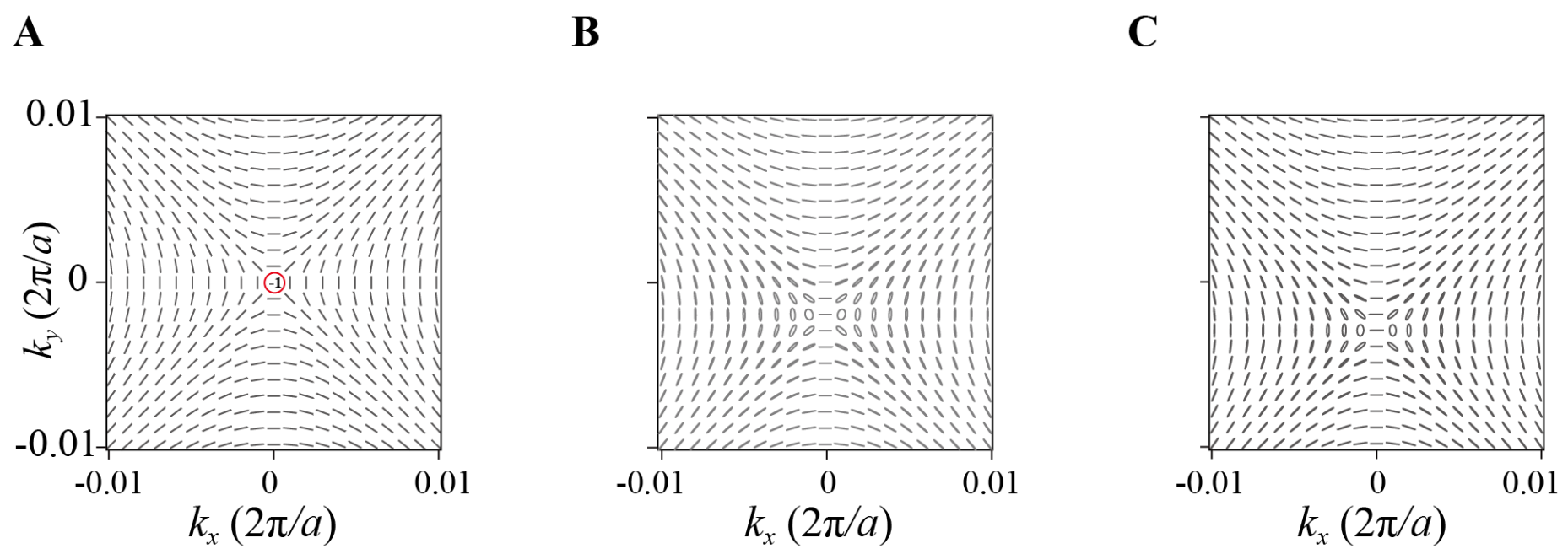

**Fig. S2 | Far-field polarization textures corresponding to Figs. 2B–D.**

(A–C) Momentum-space far-field polarization maps corresponding to the cases shown in Figs. 2B–D, respectively. The short line segments indicate the principal axes of the polarization ellipses. The polarization singularity in (A) is marked by the red circle.

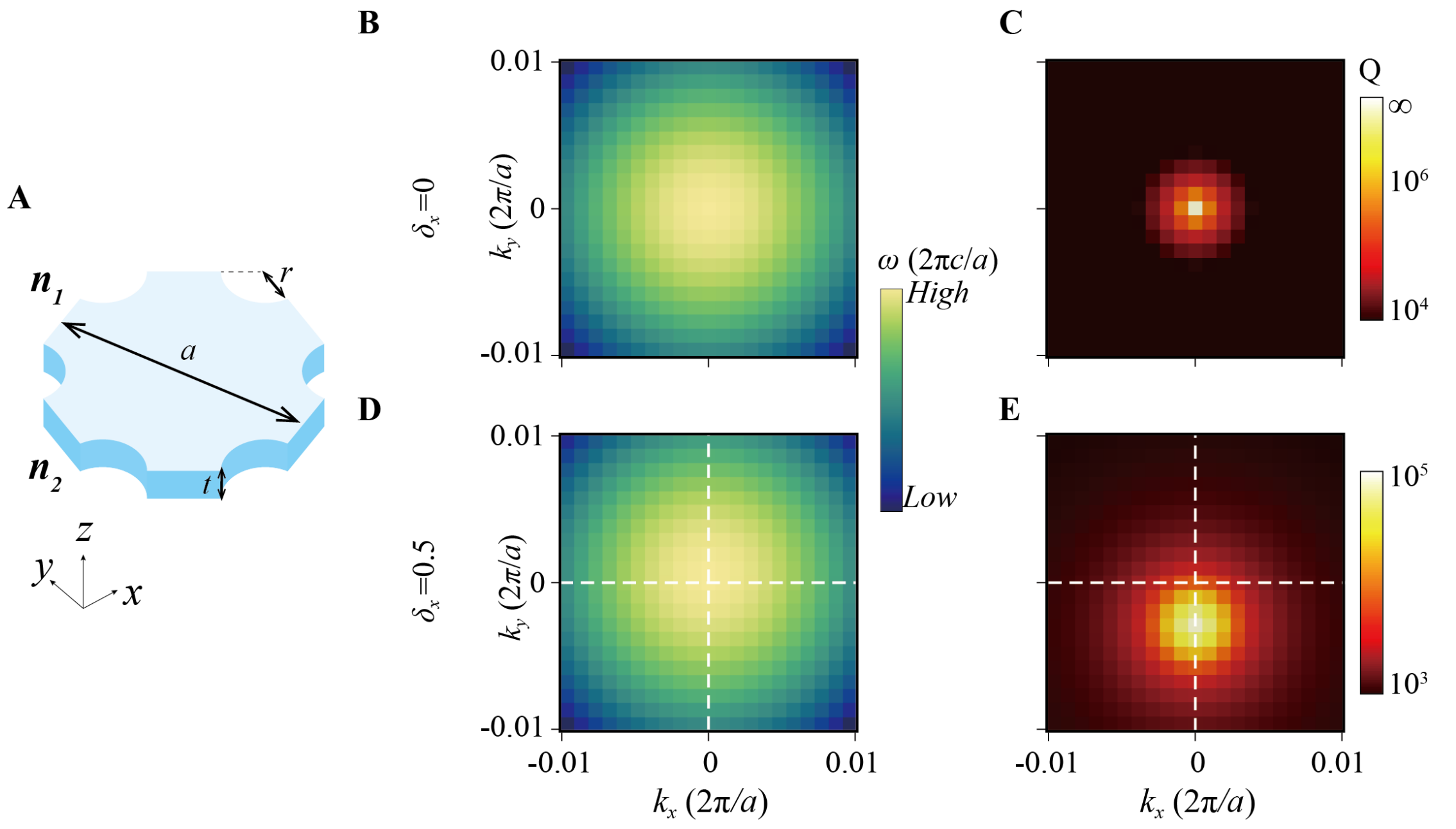

**Fig. S3 | Extension of symmetry-constrained magneto-optical beam steering to a $C_6$-symmetric photonic crystal slab.**

(A) Schematic of the $C_6$-symmetric MO-PhC slab, with lattice constant $a$ = 840nm, air-hole radius $r$ = 0.167$a$, slab thickness $t$ = 0.19$a$, and cladding and substrate refractive indices $n_1$ = 1 and $n_2$ = 1.4, respectively. (B) Momentum-space normalized-frequency distribution around the Γ point in the absence of in-plane magneto-optical coupling ($\delta_x$ = 0), showing that the band extremum remains located at Γ. (C) Corresponding Q-factor distribution, exhibiting a numerically divergent Q factor at Γ associated with the SP-BIC. (D) Normalized-frequency distribution under an in-plane magnetization along x, represented by $\delta_x$ = 0.5. The band extremum is displaced away from Γ exclusively along the $k_y$ direction, while remaining fixed at $k_x$ = 0. The white dashed lines indicate $k_x$ = 0 and $k_y$ = 0. (E) Corresponding Q-factor distribution under the same magnetic perturbation, showing that the Q-factor maximum is also displaced along $k_y$. These results demonstrate that the symmetry-constrained steering rule is not limited to the square-lattice structure considered in the main text and remains applicable to $C_6$-symmetric MO-PhC slabs that preserve the corresponding mirror symmetry.

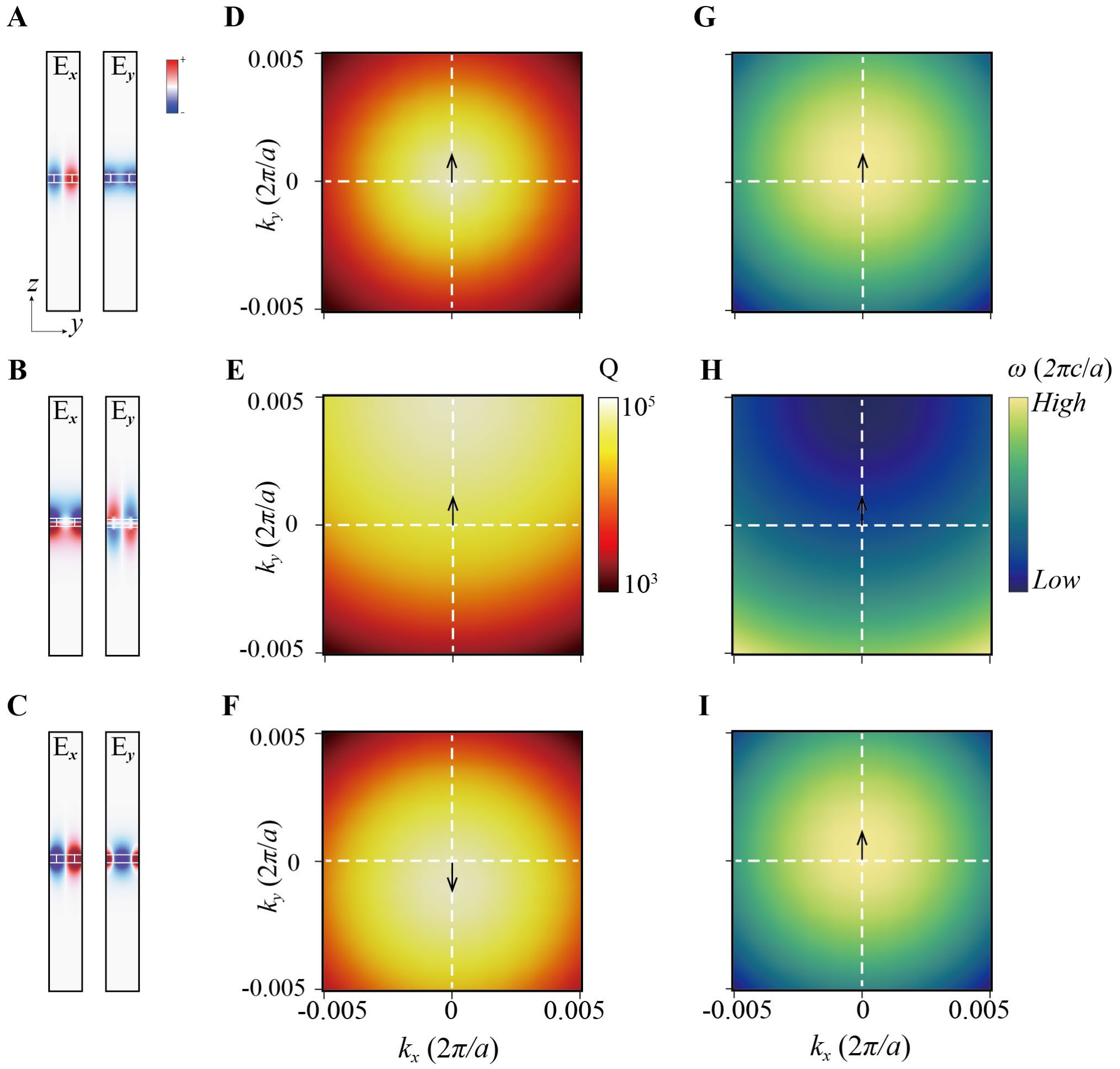

**Fig. S4 | Mode-dependent migration of the band extrema and Q-factor maxima**.

(A–C) Simulated $E_x$ and $E_y$ field distributions of three representative SP-BIC modes at Γ in the vertically symmetric and unmagnetized MO-PhC slab, showing distinct field-distribution characteristics. (D–F) Corresponding momentum-space Q-factor distributions after introducing the cladding–substrate index contrast and the in-plane magnetization Mx. (G–I) Corresponding normalized-frequency distributions, $\omega$ ($2\pi c/a$). For the modes shown in (A) and (B), the Q-factor extrema exhibit co-migration with the corresponding band extrema, whereas for the mode shown in (C), the two extrema shift in opposite directions. This comparison demonstrates that the migration of the Q-factor extrema is mode dependent and does not necessarily follow that of the band extrema.

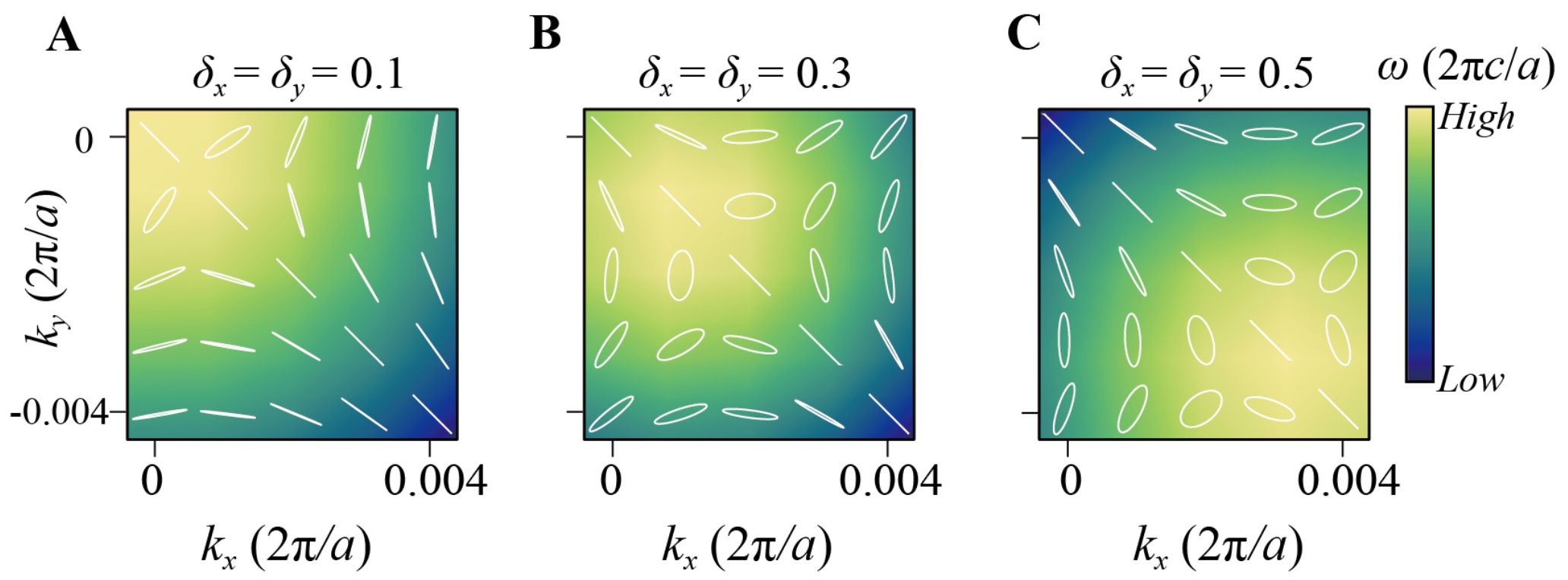

**Fig. S5 | Evolution of the local far-field polarization texture around the displaced band extremum with increasing in-plane magnetic-field strength.**

(A–C) Local far-field polarization textures calculated for $\delta_x = \delta_y = 0.1$, 0.3, and 0.5, respectively, corresponding to an in-plane magnetic field oriented at 45°. The background color represents the resonance frequency, and the white ellipses indicate the local far-field polarization states. As the field strength increases, the band extremum shifts continuously in momentum space, while the polarization at the extremum remains linear with a well-defined principal-axis orientation.

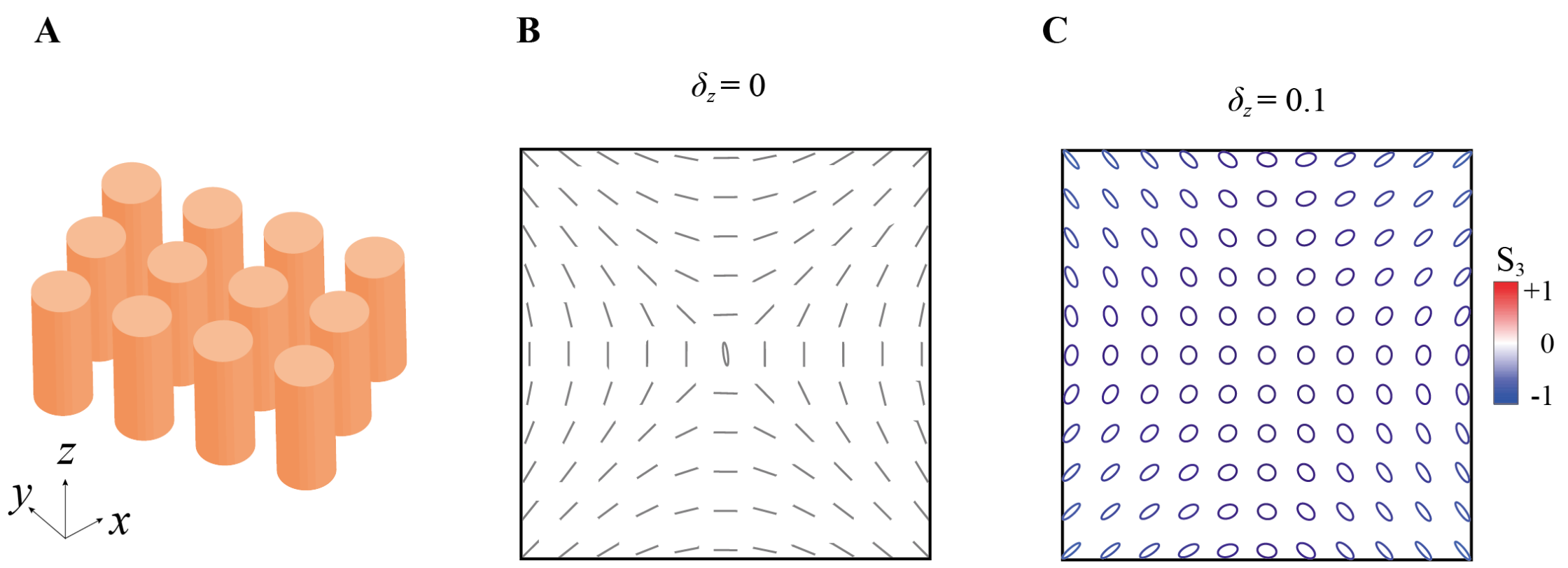

**Fig. S6 | Magneto-optical polarization response in a cylindrical-hole photonic crystal slab.**

(A) Schematic of the cylindrical-hole MO-PhC slab with a lattice constant a=1000 nm, slab thickness t=1000 nm, and hole diameter d=800 nm. (B, C) Momentum-space far-field polarization textures at the band extremum for $\delta_z = 0$ and $\delta_z = 0.1$, respectively. Compared with the square-hole structure discussed in the main text, the cylindrical-hole geometry exhibits an enhanced magnetic sensitivity due to its distinct modal field distribution, while following the same underlying magneto-optical polarization conversion mechanism.

Table S1. **Summary of the parent-mode parity and momentum-space shifts of the frequency extrema and Q-factor maxima.**

| Mode | $s$ | $\alpha$ | Predicted $k_\omega^*$ | Simulated $k_\omega^*$ | Simulated $k_Q^*$ | q |
|---|---|---|---|---|---|---|
| B | $> 0$ | $< 0$ | $+k_y$ | $+k_y$ | $+k_y$ | -1 |
| C | $< 0$ | $< 0$ | $-k_y$ | $-k_y$ | $-k_y$ | -1 |
| SA | $> 0$ | $< 0$ | $+k_y$ | $+k_y$ | $+k_y$ | +1 |
| SB | $< 0$ | $> 0$ | $+k_y$ | $+k_y$ | $+k_y$ | +1 |
| SC | $> 0$ | $< 0$ | $+k_y$ | $+k_y$ | $-k_y$ | +1 |

Modes B and C are shown in Fig. 3 of the main text, while Modes SA, SB, and SC are shown in Fig. S4. The definitions of s and $\alpha$ are given in Eqs. S11 and S9, respectively. The predicted frequency-extremum shifts are obtained from the local perturbative criterion, whereas the simulated frequency and Q-factor shifts are extracted directly from the full-wave calculations.